\documentstyle[amssymb,aps,multicol,prl]{revtex}

\draft

\begin{document}

\title{Melting of `porous' vortex matter}

\author{S. S. Banerjee$^1$, A. Soibel$^{1,*}$, Y. Myasoedov$^1$, M. Rappaport$^1$,
E. Zeldov$^1$, M. Menghini$^2$, Y. Fasano$^2$, F. de la Cruz$^2$, C. J. van der
Beek$^3$, M. Konczykowski$^3$, and T. Tamegai$^4$}

\address{$^1$Department of Condensed Matter Physics,
Weizmann Institute of Science, Rehovot 76100, Israel}
\address{$^2$Instituto Balseiro and Centro At\'{o}mico Bariloche, CNEA, Av. Bustillo
9500, Bariloche, RN, Argentina}
\address{$^3$Laboratoire des Solides Irradi\'{e}s, CNRS UMR 7642,
Ecole Polytechnique, 91128 Palaiseau, France}
\address{$^4$Department of Applied Physics, University of Tokyo, Hongo, Bunkyo-ku,
Tokyo 113-8656, and CREST Japan Science and Technology Corporation (JST), Japan}
\date{\today}
\maketitle

\begin{abstract}
Bitter decoration and magneto-optical studies reveal that in heavy-ion irradiated
superconductors, a `porous' vortex matter is formed when vortices outnumber
columnar defects (CDs). In this state ordered vortex crystallites are embedded in
the `pores' of a rigid matrix of vortices pinned on CDs. The crystallites melt
through a first-order transition while the matrix remains solid. The melting
temperature increases with density of CDs and eventually turns into a continuous
transition. At high temperatures a sharp kink in the melting line is found,
signaling an abrupt change from crystallite melting to melting of the rigid matrix.

\end{abstract}


\begin{multicols}{2}

Melting of heterogeneous systems, and in particular of nanocrystals embedded in
porous rigid matrices, is a complex process with many uncontrolled parameters.
Metal and semiconductor nanocrystals with free surfaces, for example, usually show
a decrease in their melting temperature with decreasing size \cite{couchman},
whereas nanocrystals encapsulated in a porous matrix often display an increase in
melting temperature \cite{grabaek}. Although the contribution of the different
factors is still a matter of debate, the melting process is known to depend on the
size, dimensionality, material properties of the nanocrystals and the matrix, as
well as the interface energies between the materials \cite{couchman,grabaek}. In
this work we investigate an analogous, but a more controllable composite system,
which is a `porous' vortex matter consisting of vortex nanocrystals encapsulated in
a matrix of strongly pinned vortices. As shown below, this system is present in the
commonly heavy-ion irradiated superconductors when the vortices outnumber the
columnar defects (CDs). The rigid matrix is created by vortices localized on the
network of random CDs, while the softer nanocrystals are formed within the `pores'
of this matrix by the interstitial vortices. The size of the nanocrystals can be
readily varied from several hundred down to a few vortices by changing the applied
field or the density of CDs. We find that this composite vortex matter reveals a
number of intriguing mechanisms: Similarly to the metallic nanocrystals in a
matrix, we observe for the first time a pronounced upward shift in the vortex
melting temperature $T_m$, while \emph{preserving} the first-order nature of the
transition (FOT). With increasing density of CDs, the size of the pores decreases,
resulting in a larger shift in $T_m$. We also find a critical point at which the
FOT changes into a continuous melting. Moreover, the crystallites can melt while
the matrix remains rigid. As a result, at high temperatures we find an abrupt
breakdown in the upward shift of $T_m$ and a sharp kink in the FOT line, which
apparently result from the collapse of the matrix due to vortex depinning from the
CDs.

The reported findings were obtained using Bitter decoration and differential
magneto-optical (MO) \cite{MO} techniques. High quality Bi$_2$Sr$_2$CaCu$_2$O$_8$
(BSCCO) crystals ($T_c \approx 89$ K) were covered by various patterned masks and
irradiated at GANIL by 1 GeV Pb ions with doses corresponding to matching fields of
$B_{\phi} =$ 5, 10, 20, and 50 G. Figure 1a shows schematically one of these masks
which results in the formation of CDs in BSCCO crystals only within the circular
apertures of about 90 $\mu$m diameter. This patterning allows very sensitive
simultaneous comparison of the vortex structure and the local melting processes in
adjacent irradiated and pristine regions, which is not possible by other methods.

In the absence of CDs the vortices form the Bragg glass phase \cite{GL} which has
quasi-long-range order with no topological defects, as seen in the pristine part of
the magnetic decoration image in Figs. 1b and 1c. What happens to this phase when
sparse CDs are added? The irradiated region in Fig. 1b shows that it is no longer
Bragg glass since it has significant amount of topological defects (solid circles
in Fig. 1c) and no orientational long-range order \cite{Mariela}. It is also not an
amorphous or glass phase in the usual sense, nor is it a simple polycrystal as
discussed below. In the presence of CDs the Bose glass (BG) theory
\cite{NV,LV,radzi} is usually applied, which describes the vortex matter in terms
of anisotropic \emph{homogeneously} pinned medium. Such a description is adequate
for the common situation of high irradiation doses ($B_{\phi}>B$), in which all the
vortices reside on CDs and the vortex pinning energies are comparable. In contrast,
we investigate here mainly the opposite extreme of low doses in which the vortices
greatly outnumber the CDs, i.e., $B \gg B_{\phi}$. In this case the system is
inherently \emph{heterogeneous}, consisting of two vortex populations with
\emph{well separated characteristic energies}: The vortices residing on CDs are
strongly pinned and form a rigid network or matrix, whereas the interstitial
vortices are localized by significantly weaker elastic interactions and form
relatively soft crystallites within the pores of the matrix. In the following we
refer to this state (upper parts of Figs. 1b and 1c) as porous vortex matter in
order to emphasize the important consequences of the heterogeneous structure.

Figure 2 shows several frames from a `movie' \cite{Movie} of the melting process as a function
of temperature $T$ at two fields, 30 and 60 G. Each frame is obtained by taking the
difference between the MO images at $T+0.15$ K and $T-0.15$ K and averaging a large
number of such differential images, as described previously \cite{MO}. The bright
features show the regions in the sample that undergo a FOT within the temperature
interval of 0.3 K at the indicated $T$, and the intensity of this bright
paramagnetic signal is the equilibrium magnetization step $\Delta B$ at the
transition \cite{MO,nick}. Figure 2a shows the nucleation of the liquid phase in
the form of bright inclined strips in the central pristine region which arise from
intrinsic sample disorder (see Ref. \cite{MO}). With increasing $T$ (Fig. 2b) the
liquid expands, remarkably \emph{avoiding} the irradiated apertures. In Fig. 2c the
entire central pristine part of the sample is liquid, while the apertures with
$B_\phi=20$ G are still solid. In Fig. 2d the central apertures melt at 82.45 K,
which is about 1 K above $T_m$ of the adjacent surrounding pristine regions in Fig.
2b. The apertures closer to the sample edges begin to melt in Fig. 2e. The FOT is
equally strong in the irradiated and pristine regions: The $\Delta B$ step derived
from the paramagnetic melting signal \cite{nick} is the same in Figs. 2d and 2b.
Also the width of the local melting transition is the same within our resolution,
i.e., each point in the sample melts within 0.3 K or less. This is the first direct
observation of an upward shift of the FOT by correlated disorder. Note that for
small $B_\phi$ the melting in the irradiated apertures occurs while the melting in
remote pristine regions is still in progress, and therefore the quantitative
measurement of the shift in $T_m$ has to be determined by comparing neighboring
irradiated and pristine regions and cannot be readily detected by global
techniques.

The melting process at 60 G (Fig. 2, second row) reveals two important differences.
First, the shift of the melting temperature, $\Delta T_m$, is about 4 K (difference
between Figs. 2i and 2f) which is much larger as compared to 1 K at 30 G. In Fig.
2h, for example, the entire pristine sample has melted while the irradiated
apertures are still solid. Second, the brightness of the paramagnetic melting
signal $\Delta B$ in the apertures in Fig. 2i is much lower than in the pristine
sample, and moreover, the melting in each aperture is broadened over several
frames, as exemplified in Fig. 2j. At still higher fields, above 100 G, no
paramagnetic FOT signal is detected in the irradiated apertures with $B_\phi = 20$
G.

In order to investigate the phase transition over a wider range of fields we have
used differential MO imaging in which the field is modulated by 1 G (Fig. 3)
instead of the $T$ modulation. In addition to detecting the FOT \cite{MO}, this
method provides a very sensitive measurement of the irreversibility line at a very
low effective frequency of about 0.1 Hz. Figure 3 demonstrates the determination of
the irreversibility line in the region where no FOT is present. In Fig. 3a the
entire pristine sample is in the liquid state, while the apertures are still solid.
The apertures appear black, showing that the external field modulation is shielded
due to the enhanced pinning. Upon increasing the field the black apertures
disappear (Figs. 3b and 3c) revealing the value of the local irreversibility field.

Figure 4 shows the location of the onset of the FOT for $B_{\phi}= 5$, 10, 20, and
50 G obtained by $T$ modulation (solid symbols) and of the irreversibility line
obtained by field modulation (open symbols). The solid lines are guides to the eye
for the FOT lines which terminate at the corresponding critical points. The
irreversibility data coincide with the FOT line below the critical point and
smoothly extrapolate the location of the transition line to higher fields. The
first interesting observation here is that although the structures of the porous
vortex matter and of the Bragg glass are very different (Fig. 1), their phase
diagrams for $B_{\phi}=$ 5 G in Fig. 4 are almost identical. The melting remains
FOT in most of the field range and the melting line is shifted only slightly. This
brings us to an important conclusion that the quasi-long-range order that
characterizes the Bragg glass is not an essential requirement for the existence of
a FOT \cite{colson}, and the presence of the short-range order within the
crystallites is apparently sufficient.

We can understand the upward shift in $B_m(T)$ in Fig. 4 by generalizing the
concept of the cage model. In a pure system each vortex is confined in a potential
cage arising from the elastic interactions with its neighbors. Melting occurs when
the transverse thermal fluctuations of vortices $\langle u_T \rangle$ reach a
certain fraction $c_L$ of the lattice spacing $a_0$. The pores of the matrix are
vertical cylinders which provide an additional confining cage potential to the
interstitial vortices. This enhanced rigidity reduces $\langle u_T \rangle$, and
hence stabilizes the solid crystallites within the pores, thereby shifting $B_m(T)$
upwards. The shift in $B_m(T)$ grows with $B_{\phi}$ due to the decrease in the
size of the pores. Since the pinning energy of the CDs is usually significantly
larger than the elastic energy, the melting of the crystallites may occur without
the destruction of the matrix, which remains solid up to a higher temperature as
described below. The rigid matrix may thus coexist with an interstitial liquid as
shown theoretically \cite{radzi} and in numerical simulations \cite{nandini}.

Upon increasing the density of CDs the FOT is weakened and it eventually transforms
into a continuous transition in Fig. 4. This transformation is seen more clearly by
varying the temperature along a given $B_m(T)$ line. The inset to Fig. 4, left
axis, shows the height of the equilibrium magnetization step $\Delta B$ vs. $T$ for
$B_\phi = 20$ G. As the temperature is decreased $\Delta B$ drops and eventually
vanishes at a critical point (CP) below which no discontinuity is found and the
melting becomes continuous. In addition, the width of the FOT, $\delta T_m$, shows
a significant broadening on approaching the CP (Fig. 4 inset, right axis). This
indicates that it is a true CP at which the FOT transforms into a continuous
transition as expected theoretically \cite{IW}. This CP is unlike the apparent
point-disorder-induced CP in BSCCO at lower $T$ where no appreciable broadening was
observed \cite{nature}, and where the FOT was found to be obscured by
pinning-induced irreversibility in the liquid phase \cite{Nurit}. Here, in
contrast, the liquid is fully reversible and no such experimental limitations
exist. Figure 4 shows that the CP is shifted to lower fields with $B_\phi$
\cite{Boris}. It is interesting to note that in YBCO the CP was observed to shift
upward with CDs \cite{kwok}. There, however, the CP is induced by the point
disorder in the pristine crystals and the CDs just modify slightly its position by
reducing the point-disorder-induced vortex meandering. In our case, in contrast,
the CDs apparently create a new critical point that does not exist in pristine
crystals.

We now discuss another key finding which is a kink in the $B_m(T)$ lines marked by
the arrows in Fig. 4. This kink becomes very prominent when the temperature shift
$\Delta T_m$ between the irradiated and pristine melting lines is plotted as shown
in Fig. 5. The inset displays the corresponding upward shift in field $\Delta B_m$.
In BG theory a kink in the transition line is expected to occur at $B_k\approx
B_\phi$ \cite{LV,radzi}. We argue that the kink in Fig. 5 is a new feature of
completely different nature, which reflects the collapse of the rigid matrix. The
BG kink is experimentally found to occur at $B_k/B_\phi$ of 1/6 to 1 for large
$B_\phi$, and $B_k$ usually scales linearly with $B_\phi$ \cite{kees,Tam}. The kink
in Fig. 5 occurs in the opposite regime of $B_k/B_\phi \approx 2$ to 8, and it does
not scale with $B_\phi$. At high $B_\phi$ the BG kink is a broad feature that was
found to occur only along a glass transition line \cite{kees,Tam}. In contrast,
here the kink is extremely sharp, and it is the first observation of a kink that
occurs along a FOT line. Finally, the BG kink is a transition from a strong
influence of the CDs below the kink to a much weaker effect of CDs above the kink
\cite{LV,radzi}. Here, in contrast, the situation is just the opposite, as seen in
the inset to Fig. 5. The effect of CDs is large at fields above the kink, where the
shift $\Delta B_m$ is almost constant and approximately equal to $B_\phi$, and it
collapses rapidly at $T>T_k$. This collapse occurs at $T_k \simeq 75$ K, with
little dependence on $B_\phi$. It is also interesting to note that for $B_\phi\leq
20$ G the kink occurs along the FOT line, while for $B_\phi= 50$ G it falls in the
region of continuous transition.

In Fig. 4 three regions are enumerated with respect to $B_\phi = 50$ G curve. In
region 1 the crystallites in the pores are stabilized by the rigid matrix and melt
at $B_{m}^{por}$, well above the pristine $B_m(T)$. In region 2 the pores are
liquid while the matrix remains intact, and hence $B_{m}^{por}$ reflects a melting
transition of the softer of the two substances in a heterogeneous medium. This
unconventional melting cannot be described by a single set of parameters since it
depends on the properties of both the pores and the matrix, each described by a
different set of parameters with different field and temperature dependencies. As a
result, $B_{m}^{por}(T)$ does not extrapolate to $T_c$ as commonly expected, but
rather well above it. To the best of our knowledge this is the first observation of
a melting or irreversibility line that has such an unusual property at high
temperatures. This heterogeneous process collapses sharply at $T_k$, above which
the matrix rapidly delocalizes, resulting in a homogeneous liquid in region 3. The
$B_{m}^{mtx}$ line thus describes the melting process of the matrix due to
depinning from the CDs, which leads to immediate melting of the superheated
crystallites. Thus in contrast to the common homogeneous BG description, the
$B_{m}^{por}(T)$ and $B_{m}^{mtx}(T)$ lines originate from \emph{two independent
heterogeneous processes}, melting of crystallites within a rigid matrix and the
collapse of the matrix itself, resulting in a sharp kink at the intersection point.
We expect the $B_{m}^{mtx}(T)$ line to extend also into the liquid phase
\cite{radzi}, like the dashed line in Fig. 4, separating the interstitial liquid
within a solid matrix in region 2 from the homogeneous liquid 3. Our MO
measurements cannot detect this line since both these regions are fully reversible.
In region 2, however, we expect a higher c-axis correlation than in region 3.
Recent Josephson plasma resonance studies of BSCCO with low $B_\phi$ indeed find a
recouping transition within the liquid phase at which an enhancement in the
interlayer phase coherence is observed \cite{Tam}. This extension of the
$B_{m}^{mtx}$ line should also be detectable by transport measurements since the
flux-flow resistance of the interstitial liquid should be lower than of the
homogeneous liquid \cite{radzi}. These studies will be the subject of future work.

In summary, when vortices outnumber CDs, heterogeneity rather than the average
properties of the lattice has to be taken into account for a proper description of
the structure and the thermodynamic behavior of the vortex matter. We have found
evidence for two mechanisms: melting of superheated crystallites within the pores
of a solid matrix and the destruction of the rigid matrix. The intersection point
of these two independent processes results in a sharp kink in the observed melting
line. The heterogeneous melting can be either first-order or continuous, depending
on temperature and the density of CDs. The porous vortex matter may thus provide a
tunable model system for general comprehension of melting of nanocrystals in porous
materials.

This work was supported by the Israel Science Foundation and Center of Excellence
Program, by the German-Israeli Foundation G.I.F., by the Minerva Foundation,
Germany, and by the Grant-in-Aid for Scientific Research from the Ministry of
Education, Science, Sports and Culture, Japan. EZ and FC acknowledge the support by
the Fundacion Antorchas - WIS collaboration program.


\end{multicols}

\ FIGURE CAPTIONS

Fig. 1. (a) Schematic of BSCCO crystal irradiated through a mask with an array of
circular apertures. (b) Bitter decoration image ($B=40$ G, $T=4.2$ K) showing
pristine region at the bottom (Bragg glass) and a section of the irradiated
aperture ($B_\phi = 10$ G) on top where porous vortex matter consisting of ordered
crystallites embedded in a rigid matrix is formed. (c) Corresponding locations of
six-fold (open circles) and of five- and seven-fold coordinated vortices (solid
circles) obtained by Delaunay triangulation.

Fig. 2. Melting process as a function of $T$ at fields of 30 and 60 G in BSCCO
crystal $0.65 \times 0.45 \times 0.01$ mm$^3$ using differential MO imaging with
$T$ modulation of 0.3 K. The bright regions are the areas that undergo melting
within the 0.3 K interval. (a) $-$ (c) and (f) $-$ (h): melting of the pristine
regions while the irradiated apertures are still solid. (d), (e), (i), and (j):
melting of the irradiated apertures ($B_\phi=20$ G). The area outside the crystal
is blackened. Color movies of the melting process are available at
http://www.weizmann.ac.il/home/fnsup/.

Fig. 3. Differential MO images using field modulation of 1 G in BSCCO crystal of
Fig. 2 at three fields at $T=68.4$ K. (a) All the pristine regions are liquid while
the irradiated apertures ($B_\phi=20$ G) are still solid and irreversible, and
hence appear black due to shielding. (b) Partial reversibility of the central
apertures. (c) Central apertures are liquid while the apertures closer to the edges
begin to melt.

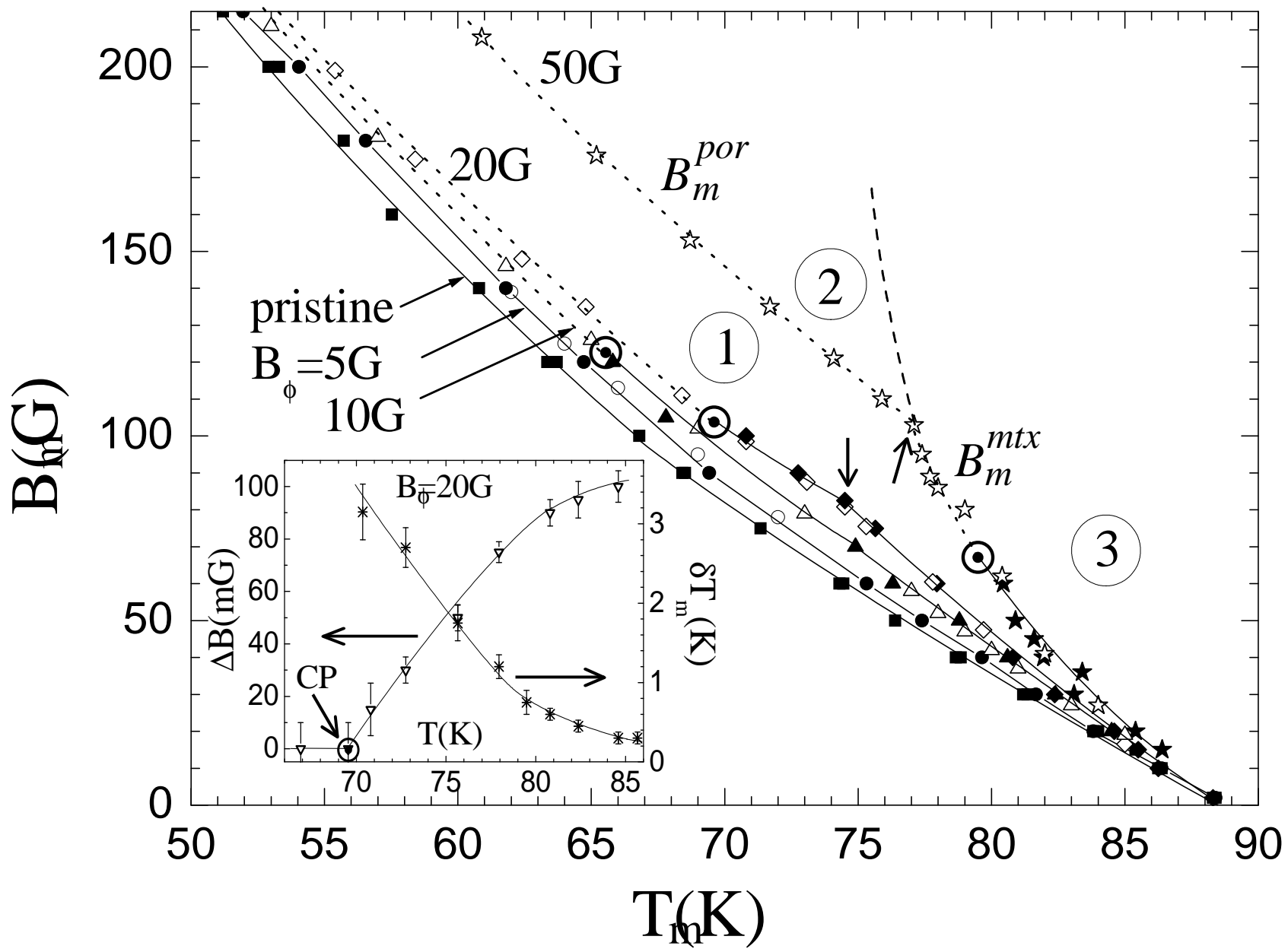
Fig. 4. The melting lines $B_m(T)$ of the pristine and irradiated regions with
indicated $B_\phi$. Solid (open) symbols are temperature (field) modulation data
showing the location of the FOT (irreversibility line). Solid (dotted) lines are
guides to the eye of the first-order (continuous) transitions that terminate at the
critical points $\bigodot$. Inset: The height of the FOT equilibrium magnetization
step $\Delta B$ which vanishes at the CP, and the local width $\delta T_m$ of the
FOT vs. $T$.

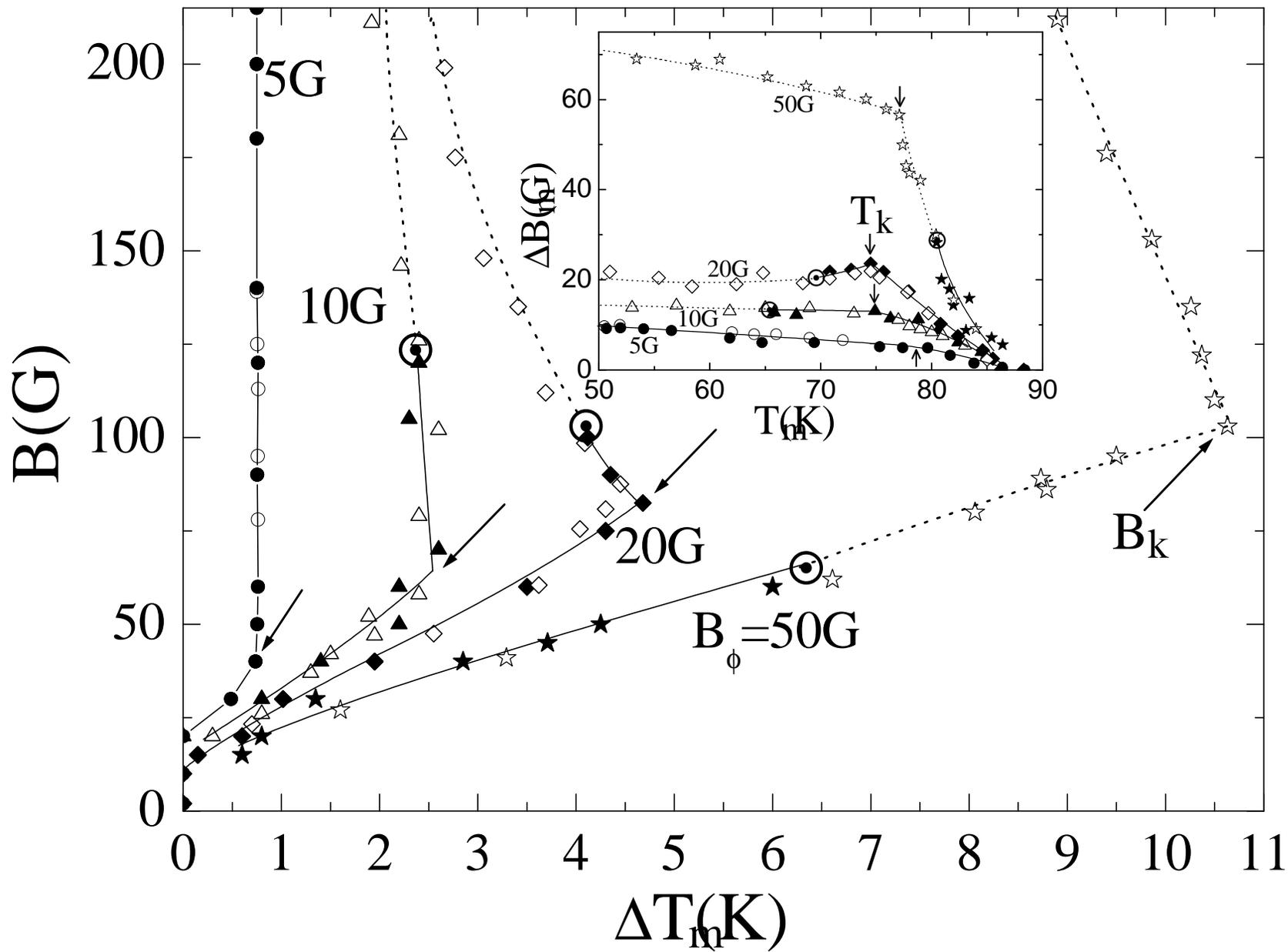
Fig. 5. The shift $\Delta T_m$ in the melting temperature for different $B_\phi$
with respect to the pristine $T_m$ at various fields $B$. Inset: The upward shift
$\Delta B_m$ in the melting field vs. $T_m$.


\end{document}